# ALIEN WAVELENGTH TECHNIQUE TO ENHANCE GARR OPTICAL NETWORK

*Paolo Bolletta[1], Massimo Carboni,[1,2], Andrea Di Peo[1], Americo Gervasi[1], Lorenzo Puccio[1], Gloria Vuagnin[1]*

[1]GARR - The Italian Research and Education Network, Rome, Italy
[2]INFN, National Laboratory of Frascati, Rome, Italy
*name.surname@garr.it

**Keywords:** Optical Networking, WAN, Alien Wavelength, Coherent, DWDM evolution

## Abstract

GARR optical network used to be composed of two separate optical network domains on its national infrastructure. With the aim to integrate these two domains and deliver high performance services all over its infrastructure, we implemented the so called alien wavelength technique, thus improving the overall efficiency of the Italian research and education network in a cost-effective way. This paper describes the activity, results, and our experience in the integration of alien wavelengths in a production environment, with a special emphasis on deployment and operational issues.

## 1. Introduction

In order to match the ever evolving needs and requirements of the Italian research and education community, we recently started an evolution process to update GARR network infrastructure and innovate the services provided. Indeed, in a global scenario characterised by an exponential growth of Internet traffic and a continuous strive for innovative solutions we started a re-definition of both our conventional network design and our common practices in device engineering [1,2].

## 2. GARR Network Overview

*2.1 GARR Community*

GARR is the Italian NREN (National Research and Education Network) connecting over 1000 sites all over Italy. Its user community is composed of universities, research institutes, research hospitals, cultural institutes, libraries, museums, and schools.

*2.2 GARR Optical Network*

GARR optical network is based on two geographically separated infrastructures. These infrastructures are based on technologies from different vendors and were deployed about four years apart. The first infrastructure deployed, is operating since 2011 in Northern and Central Italy while the second one, implemented in 2015, is operational in the South. The two infrastructures are very diverse from the technological point of view. The coexistence of such infrastructures led GARR to design an integration solution to transmit signals from the most recent technology over the older one.

From the technological point of view, the infrastructure in Northern and Central Italy employs a Huawei OptiX OSN platform. Its nodes include ROADM (Reconfigurable Optical Add and Drop Multiplexer) modules, add/drop boards able to support up to 80 channels in the C-band with a 50GHz grid and OTN (Optical Transport Network) switching matrices. The amplification is performed with EDFA (Erbium-Doped Fibre Amplifier) or Raman amplifiers, and DCMs (Dispersion Compensating Modules) are inserted on the fibre lines in order to correct the chromatic dispersion. This optical network is optimised for Intensity Modulation with Direct Detection (IM-DD) and the channels are mainly 10Gbps with few 40Gbps. Here, client services are from 1GEth up to 10GEth.

In Southern Italy, instead, the infrastructure is DCM-free and the transmission of signals is performed with coherent technology. The network is equipped with Infinera DTN-X, a platform able to transmit 500Gbps super-channels. Each of them is built on 10 optical carriers spaced at 200GHz in C-band with a 25GHz grid. Optical channels are grouped in pairs that can be enabled and managed with QPSK (Quadrature Phase Shift Keying) or BPSK (Binary Phase Shift Keying) modulation, thus allowing a flexible use of the available spectrum, and an optimal balance between reach and capacity. Client services on this part of the network range from 10GEth to 100GEth.

*2.3 Alien Wavelength Benefits for GARR*

Considered this technologically heterogeneous infrastructure, within GARR we studied a viable solution to match the high-bandwidth capacity requirements of our user community with the existing network.

As optical communications are becoming more and more spectrum efficient thanks to coherent transmissions and digital signal processing, in GARR we thought to exploit the coherent transmission infrastructure of the Southern network and to extend it to the North by means of the so-called alien wavelength (AW) technique [3,4]. Thanks to AWs, it was indeed possible to harmonise and develop GARR infrastructure by extending the 100GEth capacity already present in the Southern part of the network to the whole national infrastructure in an agile and cost-effective method.

The AW technique is a hybrid solution based on the transmission and reception of optical signals, called alien wavelengths, generated on an infrastructure, which is different from the transport one. Therefore, the transponder light is sourced from a platform and is transported in the host optical domain regardless of the Dense Wavelength Division Multiplexing (DWDM) equipment vendor. So, the AW concept disaggregates the transponder elements from the optical DWDM system.

In the specific case of GARR optical infrastructure, this method made possible the integration of the two separated optical networks in one single domain, able to homogeneously deliver the needed network services.

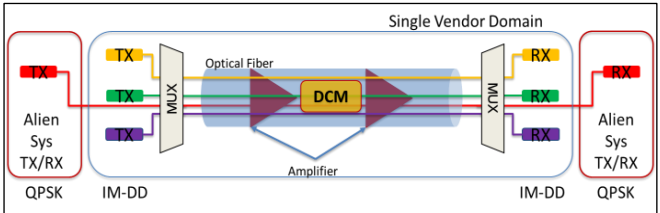

Fig. 1. Alien wavelength block diagram

Fig. 1 shows the main elements involved in AW setup as functional blocks. It illustrates how GARR deploys coherent signals over a fully compensated fibre infrastructure, by injecting signals at the service end-points and carrying the AW through the host network without regeneration and span redesign along the fibre lines. One of the main advantages of this approach is the possibility to light up network services based on different technologies without updating the geographical infrastructure and disrupting the original services, thus keeping in place legacy network services. Moreover this solution only requires new installations only at the service end-points, ensuring an effective and agile delivery.

## 3. Alien Wavelength Solution Setup

The integration between the two infrastructures (the one providing the transponders and the one providing the photonic layer) was the key issue that required careful design and operational tuning.

For this reason, we conducted a dedicated field trial to test a solution to be then implemented in the production network, following the scheme shown in Fig. 2.

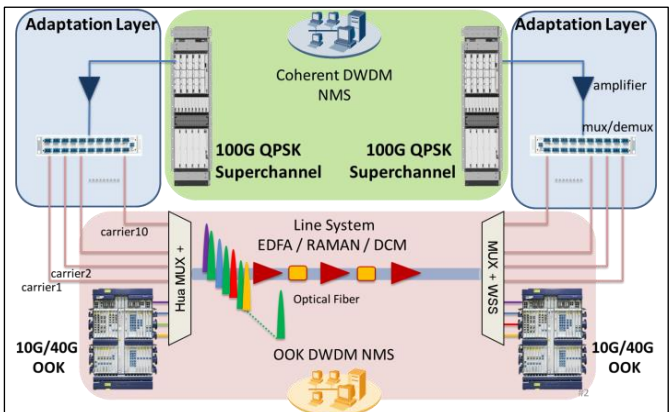

Fig. 2. Alien wavelength integration setup

---

[1] Bit Error Rate before Forward Error Correction

This designed solution can be summarised in these three main building blocks:
- *Alien wavelength domain*: The network elements generating and operating AW signals are normally installed at the service end-points. They are mainly composed of transponders mounted for example on white-boxes, DCI (Data Centre Interconnect) or OTN switches.
- *Native (host) domain*: The network elements operating and controlling the photonic layer and the fibre infrastructure. These elements compose the wide area network infrastructure handling and managing the signals across the fibres (Mux/Demux, optical amplifiers, wavelength switch, ROADMs,..). If services are already deployed on this infrastructure the respective transponders and legacy elements are part of the native domain.
- *Adaptation Layer*: The necessary element to couple the AW signals to native (host) optical domain and to adjust the injected power at the right levels.

## 4. The Alien Wavelength Experience in GARR Network

### 4.1. The Field Trial

The first aim of the field trial was to test the coexistence of native Huawei 10Gbps IM-DD optical channels with coherent alien wavelengths of a higher bit rate.

This test was useful to understand the impact of coherent alien wavelengths on the signals performance in a DCM-based optical network.

The parameters used during the performance evaluation were Q-value [6] for coherent signals and BER-PRE-FEC[1] for the others.

### 4.2. The Field Trial Results

Fig.2 shows the measurements of the average Q value of an AW super-channel versus the number of native signals sharing the same spectrum. These results pertain to a short path (345 Km long), where we studied the benchmark and the proof of concept of the solution. The first point on Fig. 2 shows the measure taken when only the alien carriers were injected in the DWDM line, with no other signals present in the spectrum. The measure was taken after performing the line and channel equalisation, and can be considered as the benchmark value for the alien performance in this specific setup. The next points of Fig. 2 were taken adding progressively one native channel (IM-DD), with (points 1-2) or without (points 3-5) guard band.

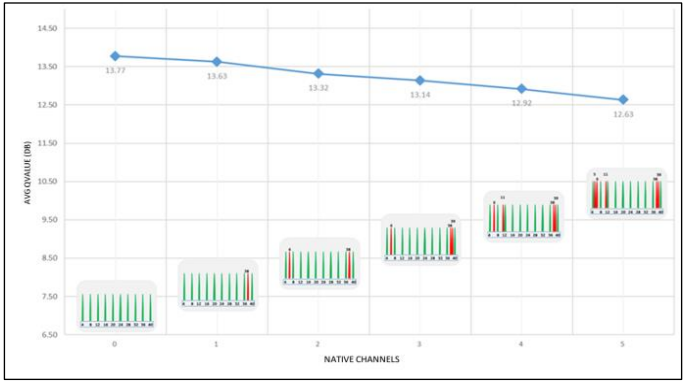

Fig. 3. Average Q-value Alien+Native AWs

The results indicate that native channels only marginally affect AW average performance, even without guard band between native and alien channels. The penalties assessment during the tests were always within 0.5dB.

It is important to remark that the Q-value of a working signal should be higher than 6.5dB, however best practices suggest to work with a value above 8.5dB for a proper link design.

We investigated the difference in performance between the QPSK and the BPSK modulation format.

Table 1: QPSK vs BPSK average Q value

|  | Average Q-value (dB | |
|---|---|---|
|  | *QPSK* | *BPSK* |
| AW | 13.77 | 16.37 |
| AW+2 native signals | 13.32 | 16.31 |
| AW+5 native signals | 12.63 | 16.15 |

Table 1 shows the Q-value achieved respectively using QPSK or BPSK modulation in three different cases:
1. only AW;
2. AW plus native signals with guard band;
3. AW plus native signals, without guard band.

As expected the signal with BPSK modulation had a better performance than QPSK and it was more robust in spite of the increase in C-band occupancy. However, the choice of using the BPSK modulation obviously reduced the super-channel capacity from 500Gbps to 250Gbps.

Moreover, the field trial was useful to understand that the coexistence in the same optical spectrum of different signals (coherent and IM-DD) did not significantly affect the IM-DD signals performance.

### 4.3. Alien wavelengths in GARR Production Network

Thanks to the field trial results, GARR designed and implemented the evolution of its network in Northern Italy using the coherent alien wavelengths in the old DCM-based infrastructure, thus creating a AW-integrated network.

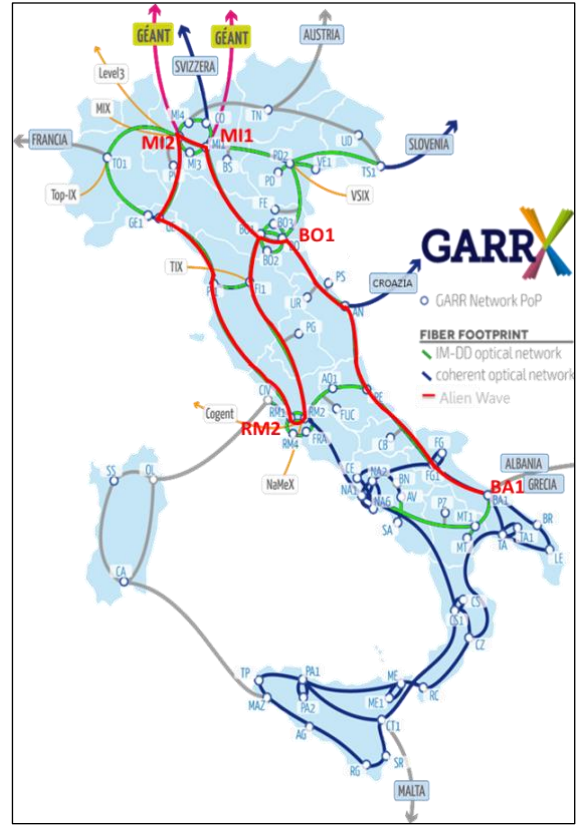

Fig. 4. GARR Network and AW deployment overview

Table 2: AW links details

| Path | *BO1-MI1* | *RM2-BO1* | *BA1-BO1* | *RM-MI2* |
|---|---|---|---|---|
| Distance (km) | 277 | 495 | 813 | 1131 |
| Attenuation (dB) | 78 | 105 | 232 | 325 |
| #OLA | 2 | 4 | 10 | 12 |
| #ROADM | 2 | 3 | 6 | 5 |
| #Raman Span | 1 | 3 | 2 | 3 |

Table n.2 shows more details on the new AW-integrated network, like distance and attenuation on the paths involved in the integration process.

Thanks to the availability of many multiplexing and adaptation elements such as WSS, MUX/DMUX and VOA, it was possible to optimise the signal performance on the paths, by fine tuning the lambdas at each transit node using the embedded signal analyser card.

### 4.4. Optimisation

During the design phase of the network it was necessary to re-engineer the frequency or the route of some production optical channels. The design study of alien wavelengths for the RM2-MI2 link has reported a very low performance index. This is because of the long link distance, about 1200Km, and the presence of adjacent IM-DD carriers at the end-points. To solve this problem and improve the performance index we chose to deploy the signal with the QPSK modulation on the AW-dedicated portion of the spectrum, instead of deploying it with the BPSK modulation on the native-AW mixed portion of the spectrum.

Indeed, the BPSK modulation option would have implied a reduction in the available capacity to 250Gbps (4.1.1), therefore we opted to dedicate a portion of the C-band to AW-coherent carriers.

However, because of the difference in the two technologies used, this dedicated C-Band portion could only support 9 carriers out of the 10 available.

As a result, through this design the available capacity moved from 250Gbps to 450Gbps on this link, in addition to the native capacity.

## 5. Results

During the field trial and after the implementation of the new design on our production network, we collected data on performance considering different distances (reach). The following graph (Fig.3) summarises these measurements, by comparing the performance of AW in their native environment in Southern Italy (diamond shapes) versus their performance in the trial paths (square shapes) and in the production environment (triangle shapes).

The vertical bars on the production data indicate the highest and lowest performance measured in Q-value among the carriers, thus illustrating the total range of values collected in the production environment.

It is important to underline that the Q-value measured at 11.44dB was obtained in the dedicated portion of the optical spectrum.

We can observe that all production signals have an acceptable performance margin, above the Q-value acceptance threshold.

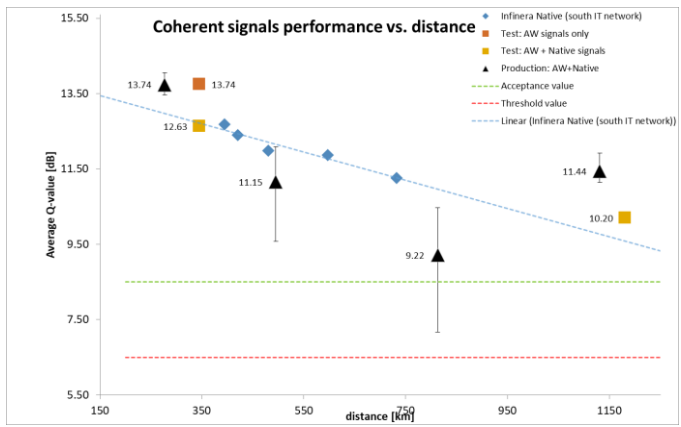

Fig. 5. Coherent signals performance vs. distance

## 6. CONCLUSION

Thanks to the encouraging results in the field trial and in the production environment, GARR managed to match the requirements of its users community to provide 100GEth client services on the main backbone nodes of its national transport infrastructure. Also, in this way GARR managed to have an integrated high-capacity network bridging the different technologies of its optical infrastructure. From the user community perspective, the most important effect was an increased availability of bandwidth.

For the future, this enriching experience will be the launch pad towards the forthcoming evolution of GARR network [5].

## Aknowledgments

We thank Elis Bertazzon for her assistance with the technical editing process of the present document.